# Electromagnetic Near-Field Mutual Coupling Suppression with Active Janus Sources

Bo Xue[1], Kayode Adedotun Oyesina[1] and Alex M. H. Wong[1,2] *

[1]*Department of Electrical Engineering, City University of Hong Kong, Tat Chee Avenue, Kowloon, Hong Kong SAR, China*
[2]*State Key Laboratory of Terahertz and Millimeter Waves, City University of Hong Kong, Tat Chee Avenue, Kowloon, Hong Kong SAR, China*

*Correspondence should be addressed to:

Alex M. H. Wong (alex.mh.wong@cityu.edu.hk)

## Abstract

Electric dipoles and magnetic dipoles are the most fundamental particles in electromagnetic theory. Huygens and Janus sources, formed by the orthogonal combination of electric and magnetic dipoles, both show good directionality in the near field. Although the Huygens source has been widely used in antennas and metasurfaces, the applications of Janus source are heretofore limited. In this paper we report the first physical construction of an active Janus source. Through full-wave simulations within the PPW environment, we show that our source achieves the directional electromagnetic near-field and quasi-isotropic far-field requisite of the Janus source. Using this fact, we demonstrate that two active Janus sources in close proximity (about 0.10 to 0.25 wavelengths) achieve a near 1000-fold reduced mutual coupling compared to electric dipole sources. The achievement of strong mutual coupling suppression and quasi-isotropic radiation make the Janus source an ideal candidate for consideration in future compact MIMO communication systems.

## I. Introduction

Electric and magnetic dipoles, the fundamental sources of electromagnetic waves, by themselves produce omnidirectional radiation in the plane perpendicular to the dipole. However, their simple combinations quickly result in intricate directionalities of wide-ranging interest to physics, photonics and engineering. It is well known that a pair of co-located electric and magnetic dipoles, equal in radiation strength and phase and orthogonally directed in space, form a Huygens source which produces unidirectional radiation. This radiation characteristic was first described by Fresnel in his introduction of the obliquity factor [1]. Though at the time the reason for its occurrence was unknown, its introduction reconciles Huygens' wave theory of light with experimental observations. Through later developments from Kirchhoff [2],

Love [3] and Schelkunoff [4] we now understand that the obliquity factor occurs as a consequence of directive radiation of the Huygens sources that one obtains when one replaces a propagating plane wave with its secondary (or equivalent) sources along a constant phase plane. The unidirectional property of the Huygens source has been extensively studied, and passive and active Huygens sources find applications in various fields. Passive Huygens sources are used to attain flexible control of coupling direction to achieve directionality modulation [5–7]; Huygens metasurfaces are proposed to achieve flexible modulation of electromagnetic waves with high efficiency [8–11]. In the field of antennas, the magnetoelectric (ME) dipole [12,13] and the electrically small Huygens source antennas [14,15] use a combination of electric and magnetic dipoles to construct high gain antennas leveraging the far-field directionality of Huygens source.

While the Huygens source achieves directionality in the far-field, a recently proposed Janus source achieves directionality in the near-field. The Janus source features an electric and magnetic dipole pair which are co-located, orthogonally directed and with equal radiation strength, just like the Huygens source. However, in departure from the Huygens source, the electric and magnetic dipoles are shifted in phase by 90 degrees. Picardi et al. investigates parity and time-reversal of the circular, Huygens and Janus dipoles, and find that they possess distinct and unique directionalities in the near-field [5]. Recent works show important applications Janus sources have in near-field power coupling scenarios [5,16–28]. However, these works either treat the Janus source from a theoretical perspective with an ideal current doublet [5,16,21,24,26], or employ a passive source, such as a nanosphere [17], dielectric cylinder [27] or helix [25], which scatters a very small fraction of an incident wave in a manner similar to a Janus source. Hence passive Janus sources are excited with limited power efficiency, which somewhat limits the possibility of their practical application. Further, requirement of an incident illumination, and the background EM field and scattering caused by it, forms spurious fields that are often undesirable. An active Janus source, which generates the desired electric and magnetic currents in the correct amplitude, phase and orientation, would be very desirable toward the experimental investigation and practical application of Janus sources. In addition, the active Janus source is more efficient and the modulation is more flexible compared to the passive Janus source.

One potential application of the Janus source is in closely spaced MIMO antenna systems. Having multiple closely spaced antennas can increase the capacity of the communication system, but the electromagnetic near-fields of these antennas often interfere with one another, resulting in performance degradation of various kinds. Complicated decoupling networks are designed to suppress mutual coupling with varying degrees of success [29–33]. Pattern diversity method has also been used by employing antennas with different radiation patterns [34–38]. In this manner, mutual coupling can be reduced, but the communication performance may have angular variations. For example, for a MIMO antenna system transmitting two channels of information with different antenna patterns, the corresponding receivers may have dominant reception of either Channel A or Channel B, depending on their angular location with respect to the transmitter. Since the Janus source features near-field directionality while it possesses a near-isotropic far-field pattern, it has potential to suppress mutual coupling in a compact MIMO antenna system without compromising the radiation properties of individual antennas.

In this paper, we address the aforementioned scenarios by demonstrating (i) a practical active Janus source, and (ii) the dramatic suppression of mutual coupling using active Janus sources. We experimentally report, to the best of our knowledge, the first realistic active Janus source, which we achieve by properly tuning the current running through two filaments placed within a parallel-plate metallic waveguide. This source can be generated by feeding two monopole antennas with a relatively simple microwave circuit. We then demonstrate that, when two such sources are placed within close proximity (from 0.08 to 0.24 wavelengths), the mutual coupling between them is reduced around 1000-fold compared to two dipoles of similar proximity. Our findings show that the Janus source can be excited in an active manner, with high power efficiency and without affecting the neighbouring sources, and that they can be used to suppress mutual coupling between two radiators in-close proximity to one-another. The work paves way to power-efficient active directional near-field power-coupling devices and opens doors to a new kind of compact MIMO antenna systems.

## II. Active Janus Source

We introduce a 2D environment in which we investigate and demonstrate the active Janus source. Figure 1(a) shows a parallel plate waveguide (PPW) environment with two perfect conductors normal to the z-direction. When separated at a distance $h < \lambda/2$, where $\lambda$ is the electromagnetic wavelength, this environment only supports modes with non-zero $\{H_x, H_y, E_z\}$. All fields remain invariant in the z-direction, hence the effective 2D environment is constructed. Performing our investigation in a 2D (PPW) environment helps simplify our analysis. The conclusion from this work is nonetheless directly applicable to a 3D environment.

We have shown in earlier works [11,28,39,40] that the twin current filament (TCF) source shown in the inset of Fig. 1(a) can generate co-located, spatially orthogonal electric and magnetic current pairs in this environment. Essentially, when the current filaments are separated by a distance much smaller than the wavelength ($w \ll \lambda$), they form an effective electric current $I_E$ in the z-direction that is the sum of the currents flowing in both filaments. Conversely, an effective current loop $I_M$ is formed with the difference of currents flowing in the filaments, which functions as an effective magnetic current in the yz-plane. Mathematically,

$$I_E = I_b + I_a$$
$$I_M = \frac{1}{2}(I_b - I_a) \tag{1}$$

where $I_a$ and $I_b$ are electric currents flowing through the two filaments.

Following Eq. (1), a pure electric current source can be generated with $I_a = I_b$, while a pure magnetic current source can be generated with $I_a = -I_b$. More complicated superpositions can be achieved by properly tuning $I_a$ and $I_b$. Particularly, the Huygens and Janus sources can be generated by requiring that the resultant magnetic and electric currents $\boldsymbol{M}$ and $\boldsymbol{J}$ possess the same radiation strength and are either in phase (Huygens source) or phase-shifted by 90° (Janus source). In Appendix A we overview the derivation relating the electric and magnetic currents to the infinitesimal electric dipole and infinitesimal electric loop. From this, we can find the superposition of the electric dipole and electric loop currents that will give rise to the appropriate J and M which forms a Huygens and/or a Janus source. Applying Eq. (1) to Eqs. (S3) and (S4) from Appendix A, we find that the Huygens and Janus sources can be generated when

$$\frac{I_{bH}}{I_{aH}} = -\frac{h + j2kA}{h - j2kA} = -\frac{1 + j4\pi(w/\lambda)}{1 - j4\pi(w/\lambda)} \tag{2}$$

for the Huygens source, and

$$\frac{I_{bJ}}{I_{aJ}} = -\frac{h + 2kA}{h - 2kA} = -\frac{1 + 4\pi(w/\lambda)}{1 - 4\pi(w/\lambda)} \tag{3}$$

for the Janus source.

We introduce a single excitation current $I_0$, for which

$$|I_0|^2 = |I_a|^2 + |I_b|^2 \,;\; \Delta\psi = \psi_b - \psi_a \tag{4}$$

where the $\psi_n$ is the phase of current excitation $I_n$ ($n$ = a, b). Using Eq. (4) we rewrite the excitation conditions for the electric, Huygens and Janus dipoles as

$$I_{aE} = I_{bE} = \sqrt{2}I_0; \;\; \Delta\psi_E = 0; \tag{5}$$

for the electric dipole,

$$I_{aH} = \sqrt{2}I_0 e^{j\,\text{atan}(\xi^{-1})}; \quad I_{bH} = \sqrt{2}I_0 e^{-j\,\text{atan}(\xi^{-1})}; \quad \Delta\psi_H = 2\,\text{atan}(\xi^{-1}); \tag{6}$$

for the Huygens source, and

$$I_{aJ} = \frac{1-\xi}{\sqrt{2(1+\xi^2)}} I_0; \quad I_{bJ} = -\frac{1+\xi}{\sqrt{2(1+\xi^2)}} I_0; \quad \Delta\psi_J = \pi, \quad \text{where } \xi = \frac{4\pi w}{\lambda} \tag{7}$$

for the Janus source. Thus properly tuning $I_0$ and $\Delta\psi$, we can actively excite the electric / magnetic dipole, Huygens and Janus sources within a 2D PPW environment.

We emphasize that the TCF is a realistic source that can be directly implemented with a microwave circuit. In [11] we experimentally demonstrated a simpler version of the TCF as a Huygens source along a metallic boundary. For the TCF in this work, we will excite the filaments by connecting them to co-axial cables from one side of the PPW. The proper current feeds can be straightforwardly generated by using an asymmetric power splitter and a meander line phase-shifter. We will report more details in the experiment section of this paper.

We proceed to investigate the TCF using full-wave simulations with Ansys HFSS. We adopt a simulation frequency of 800 MHz ($\lambda_0 = 375$ mm). The key TCF dimensions are $h = 0.12\lambda_0$, and $w = 0.04\lambda_0$. The PPW has a size of $l = 4\lambda_0 = 1500$ mm; it is terminated by a radiation boundary to simulate infinite wave propagation. We place the TCF at the middle of the PPW and observe the spatial field distributions inside the PPW.

As shown in Fig. 1(a), two currents $I_a$ and $I_b$ in z-direction comprise the active sources in this PPW environment. The current sources are excited as lumped ports where the port impedance is tuned to minimize reflection for a single TCF. This ensures the power delivered to the filament is radiated into the PPW. After we have optimized the port impedance, we excite the ports according to Eqs. (5-7) to generate electric dipole, Huygens source and Janus source excitations respectively. Figures 1(b, d, f) show the simulated |**E**| field distributions associated with these three excitations. To compare with theory, we also show, in Fig. 1(c, e, g), the simulated and calculated radiation patterns along the principal (xy) plane for an electric dipole, a Huygens source and a Janus source. Here the blue curves illustrate the radiation patterns calculated from the superposition of infinitesimal electric and magnetic dipoles with the requisite amplitudes and phases, within a 3D free-space environment. The dipole orientations of each case are labelled onto the radiation patterns. The red curves are the simulated "radiation" patterns (i.e. waves travelling outwards on the xy-plane within the infinite PPW waveguide). We can see that the simulated results are consistent with the calculated results. As expected, the electric dipole radiates omnidirectionally on the xy-plane, the Huygens source shows unidirectional radiation into the +y-direction, and the Janus source yields a quasi-isotropic radiation with about 3dB variation across the angles, achieving maximum radiation in the ±y-directions and minimum radiation (with roughly 70% field strength) in the ±x-directions. Slight disagreements for the radiation of the Huygens source can be observed in Fig. 1(e): this is understandable as the fields calculation is performed for a 3D infinitesimal source and while simulation is performed for a 2D environment. Notwithstanding this, an agreement in the general radiation trend is observed. These results, particularly the agreement between Fig. 1(f, g), validate our generation of the active Janus source using the TCF structure in this 2D PPW environment. Further, it shows that the 2D active sources demonstrated hereby attain very similar radiation characteristics as their 3D counterparts do in free space, along the principal plane. This makes our findings here directly applicable to 3D active sources in a free-space environment.

Moreover, we investigate the near-field coupling properties of the active Janus source [28]. To do this, we simulate the geometry of Fig. 2(a), which is similar to the waveguide coupling topology in [5]. Two dielectric waveguides are placed at a sub-wavelength distance $d_1$ away from the active source. The dielectric waveguides are symmetric with respect to the origin along the y-axis in the case of Huygens source and along the x-axis in the case of Janus source. Depending on the source excitation characteristics, the generated electromagnetic near-fields couple to the waveguides in with strong directional preference. Figures 2(b-c) show the directionality of the excited waveguide modes when the Huygens source is excited to direct radiation in the +y and -y directions respectively. Figures 2(d-e) show the directionality of the excited waveguide modes when the Janus source is excited to direct near-field coupling to the +y and -y directions respectively. The near-field directionality exhibited by the Huygens and Janus sources agree with those produced with ideal sources in [5]. The achievement of both the far-field radiation pattern and the near-field directionality strongly indicates the successful construction of an active Janus source within the 2D PPW environment.

## III. Near-Field Mutual Coupling Suppression with Active Janus Sources

Having successfully generated the Janus source, we leverage its distinctive near-field directionality to reduce coupling between two closely spaced sources. First, we construct two active sources shown in Fig. 3(a) based on the TCF model introduced in Section II. All dimensions are the same as the TCF of Fig. 1(a). To investigate the mutual coupling between two such sources, we separate them by a center-to-center distance $d$ along the y-direction. Specifically, we vary $d$ from $0.08\lambda_0$ to $0.24\lambda_0$ at an interval of $0.04\lambda_0$ to investigate the mutual coupling in the near-field. The excitation of the pair of active sources $\{I_{1aN}, I_{1bN}\}$ and $\{I_{2aN}, I_{2bN}\}$ are found from Eqs. (5) – (7). Here $N = E, H, J$ represent the electric dipole, Huygens source and Janus source respectively, and the subscripts 1 and 2 denote the first and second source. It is particularly important to note that in order to achieve coupling suppression between the two sources for the Huygens and Janus sources, one should orient the sources such that their near-field directionalities are in opposite directions. For instance, if Source 1 points in the $-y$ direction, Source 2 should point in the $+y$ direction.

To characterize the mutual coupling between Sources 1 and 2, we design an RF circuit, as shown in Fig. S1(a), that feeds both sources with a single port. The circuit contains a Wilkinson power splitter that splits the incident wave into the correct amplitude ratio of $|I_a|$ and $|I_b|$. Thereafter, the branch with current amplitude $I_b$ is meandered to achieve the required phase shift $\Delta\psi$. This simple feed network is quite common in microwave circuit design. Here it is theoretically modelled to obtain its scattering matrix $[S_{WPS}]$ (see Appendix B). Two such scattering matrices connected to Sources 1 and 2, as well as scattering matrix $[S_{PPW}]$ obtained from full-wave simulation representing the port relationships when the two TCF sources are placed within the PPW environment, are combined using microwave network theory to obtain the simulated mutual coupling level. Appendix B explains this procedure in detail. The combined system can now be described by its own scattering matrix. In particular, the parameter $S_{total,21}$ denotes the ratio of the output wave at Port 2 to the input wave at Port 1, which thus characterizes the mutual coupling between the two sources.

Using this formulation, we investigate the mutual coupling within for two TCFs separated at distance $d$ within the PPW environment. A strength of our simulation method is that it requires only one full-wave simulation for each separation distance. We sweep the phase difference $\Delta\psi$ and the amplitude $|I_a|$ (the corresponding $|I_b|$ is thus $|I_b| = \sqrt{|I_0|^2 - |I_a|^2}$) by modifying $[S_{Feed}]$ to observe the mutual coupling for all possible combinations of $\{|I_a|, |I_b|, \Delta\psi\}$, including the electric dipole ($\{\sqrt{2}I_0, \sqrt{2}I_0, 0°\}$), Huygens source ($\{\sqrt{2}I_0, \sqrt{2}I_0, 165°\}$) and Janus source ($\{0.79I_0, 0.61I_0, 180°\}$). The excitation parameters for all sources are found from Eqs. (5) - (7). As an example, Figure 3(b) shows the variation in mutual coupling with varying $\Delta\psi$ and $|I_a|$, for $d = 0.08\lambda_0$. The conditions for the electric, Huygens and Janus sources are labelled. It can be observed that the region with low coupling coefficients are concentrated in the lower right region due to the presence of two mutual coupling nulls. Both the Huygens source (denoted by a circle marker) and the Janus source (denoted by a diamond marker) are in within the vicinity of this low-coupling region, hence they experience much weaker coupling compared to a pair of dipole sources. At this distance the mutual coupling between the Huygens and Janus source pairs are reduced by 19.5 dB and 25.44 dB respectively compared to the dipole source pair.

Likewise, Figures 3(c-f) show the mutual coupling for all source pairs for distances from $0.08\lambda_0$ to $0.24\lambda_0$ at intervals of $0.04\lambda_0$. We observe that as $d$ increases, the weakly coupled region remains within the lower right corner of the 2D colour map, and gradually moves closer to the location of the Huygens source. Across all distances, the Huygens and Janus source are always in the weakly coupled region and the coupling suppression is significantly better than that of the electric dipole. Table I plots the coupling coefficients of the three different sources as a function of $d$. In terms of the simulated results, the coupling between two Huygens source pair is suppressed by more than 30 dB (1000 ×) compared the dipole source pair for $d \geq 0.16\lambda_0$. The coupling between the Janus source pair is suppressed by around 29 dB (794 ×) compared the dipole source pair for $d \geq 0.16\lambda_0$. In both cases, dramatic mutual coupling reduction is achieved using Huygens and Janus source pairs. It can be seen that the Huygens source pair achieves stronger suppression with increasing separation distance, while the Janus source pair seems to achieve a steady suppression level from $d = 0.12\lambda_0$ to $d = 0.24\lambda_0$. Beyond this distance, the trend continues for the Huygens source pair while the Janus source pair ceases to produce strong coupling suppression. This can be explained by the fact that the Huygens source exhibits both near- and far-field directionality while the Janus source exhibits near-field directionality but is quasi-isotropic in the far-field.

## IV. Experimental Demonstration

We proceed to experimental verification. A diagram of the experimental setup is shown in Fig. 4. Two 1200mm × 1200mm square aluminum plates separated at a subwavelength height $h = 45$mm form the top and bottom plates of the PPW. Microwave absorbers are placed at the PPW boundary to prevent reflection from the edge of the PPW. The active sources are built using monopole antennas as the driven elements. An array of holes was drilled in both aluminum plates for the insertion of the monopoles. The corresponding passive lossless microwave feed circuits for each of the three sources — electric dipole, Huygens source and Janus source are designed to achieve the specific excitation weights necessary to implement the sources according to Eqs. (5) - (7). Port 1 of a Keysight E5071C Vector Network Analyzer (VNA) provides the input power to microwave feed network 1 with its two output ports exciting source 1 (electric dipole, Huygens, Janus) at prescribed excitation amplitudes and phases.

Similarly, the VNA's Port 2 provides the input power to microwave feed network 2, with its two output ports exciting the dipole source 2 (dipole, Huygens, Janus) as appropriate (refer to Fig. 5(a)). The coupling coefficients (S21) are then measured by the VNA.

Figures 5(b-f) plot the measured coupling coefficient as functions of frequency for different values of $d$. Table I shows the measured coupling coefficients for 800 MHz. The measured results show good agreement with the simulation. For all separation distances, mutual coupling is clearly suppressed for the Huygens and Janus sources pairs compared to the electric dipole source pair: Janus source pair achieves a suppression level exceeding 25dB ($> 316 \times$) for all distances measured, while the Huygens source pair achieves a suppression level of 20-25 dB (100 to 316 $\times$) for distances up to $0.20\lambda_0$, and a suppression level approaching 30dB (for 1000 $\times$) for $d = 0.24\lambda_0$. The mutually coupled power for Janus source pair is lower than Huygens source pair for $d = 0.08\lambda_0$ and $d = 0.12\lambda_0$, but at higher separating distances the Huygens source pair shows a stronger suppression in mutual coupling, consistent with observations from full-wave simulation. It is therefore clear that both Huygens and Janus sources can serve as good candidates in applications requiring low inter-element coupling; further, the Janus source's ability to achieve strongly suppressed mutual coupling while featuring a quasi-isotropic far-field gives points to intriguing application potential as antennas for compact MIMO systems.

## V. Conclusion

In this paper we have demonstrated the first realistic active Janus source and reported near-field mutual coupling suppression between a pair of such active Janus sources. We design a simple, realistic active Janus source based on a succinct PPW structure. The presented model can function as all three sets of fundamental dipolar sources (electric / magnetic dipole, Huygens source, and Janus source) under the right excitation, thereby allowing us to investigate the near-field mutual coupling when these fundamental source pairs are in close proximity to each other. The simulation and experiment results show a many-fold (up to 1000 times) suppression of coupling in the Huygens and Janus source pairs when compared to a pair of electric dipoles at similar separation from each other. The findings in this work open door to the exploration of these directional sources in densely packed systems where coupling between cells has imposed limitations. More interestingly, since the Janus source is quasi-isotropic in the far-field, the significant near-field coupling suppression positions it as a good candidate for future compact MIMO antennas design.

# Appendixes

# Appendix A: Electromagnetic Field Analysis of Huygens and Janus Sources

We consider electric-magnetic dipole pair where the electric dipole with current $I_E$ and an infinitesimal length $l$ is oriented along the z-direction, and the magnetic dipole, synthesized with a loop with electric current $I_M$, occupies an infinitesimal planar area $A$, with the surface normally pointing along the x-direction.

Using field relationships for electric and magnetic Hertzian (infinitesimal) dipoles [41], the electric field and magnetic fields due to the two currents, on the xy-plane, can be written as:

$$E_\theta = jk\eta I_E \left(\frac{l}{4\pi r}\right)\left[1 + \frac{1}{jkr} - \frac{1}{(kr)^2}\right]e^{-jkr} - \eta k^2 I_M \left(\frac{A\sin\phi}{4\pi r}\right)\left[1 + \frac{1}{jkr}\right]e^{-jkr};$$

$$H_r = jkI_M \left(\frac{A\cos\phi}{2\pi r^2}\right)\left[1 + \frac{1}{jkr}\right]e^{-jkr}; \quad (S1)$$

$$H_\phi = jkI_E \left(\frac{l}{4\pi r}\right)\left[1 + \frac{1}{jkr}\right]e^{-jkr} - k^2 I_M \left(\frac{A\sin\phi}{4\pi r}\right)\left[1 + \frac{1}{jkr} - \frac{1}{(kr)^2}\right]e^{-jkr}.$$

All other electromagnetic field components vanish. Here $k$ and $\eta$ represent the wavenumber and intrinsic impedance of the host medium and $\{r, \theta, \phi\}$ represent the spherical coordinates of the observation point with respect to the sources (which are located at the origin). As usual, $\theta = 0°$ is along the +z-axis and $\phi = 0°$ is along the +x-axis.

We first generate, with these two dipoles, the Huygens source with power transfer to the +y-direction ($\phi = 90°$). To achieve this, the two terms for electric (and magnetic) far-fields must be of the same magnitude and must constructively interfere in the +y-direction. The far-field of Eq. (S1) can be approximated as

$$E_\theta = jk\eta I_E \left(\frac{l}{4\pi r}\right)e^{-jkr} - \eta k^2 I_M \left(\frac{A\sin\phi}{4\pi r}\right)e^{-jkr};$$

$$H_\phi = jkI_E \left(\frac{l}{4\pi r}\right)e^{-jkr} - k^2 I_M \left(\frac{A\sin\phi}{4\pi r}\right)e^{-jkr}. \quad (S2)$$

Here we have kept only the highest order term of $r$. We now require the two additive terms of $E_\theta$ and $H_\phi$ to have the same complex amplitude. This is done by stipulating

$$jk\eta I_E \left(\frac{l}{4\pi r}\right) = -\eta k^2 I_M \left(\frac{A}{4\pi r}\right) = E_{H0} = \eta H_{H0},$$

$$\Rightarrow \frac{I_E}{I_M} = j\frac{kA}{l}. \quad (S3)$$

We proceed to investigate, in similar fashion, the fields emanated by a Janus source. We employ a similar substitution as Eq. (S3) but include a $\frac{\pi}{2}$ phase separation between the electric and magnetic currents, hence

$$jk\eta I_E \left(\frac{l}{4\pi r}\right) = -j\left(-\eta k^2 I_M \left(\frac{A}{4\pi r}\right)\right) = E_{J0} = \eta H_{J0}$$
$$\Rightarrow \frac{I_E}{I_M} = \frac{kA}{l} \qquad (S4)$$

We can observe that the magnitude between $I_E$ and $I_M$ is determined by electric dipole's length, electric loop's area as well as wavenumber. Furthermore, the phase difference between $I_E$ and $I_M$ is 90° for Huygens source and 0° for Janus source. This is because the magnetic dipole has 90-degree phase shifted with the electric loop current.

## Appendix B: Network Analysis for Near-field Mutual Coupling System

Figure 3(a) shows our proposed configuration involving a pair of oppositely excited Janus sources, separated (center-to-center) at a distance $d$. The source pair is located in the middle of a PPW $3\lambda_0 \times 3\lambda_0$ in size. The waveguide extremities are terminated by absorbing boundaries. Full-wave electromagnetic simulations are performed to obtain the S-matrix $[S_{PPW}]$ describing the relationship of the 4 ports: $\{1a, 1b, 2a, 2b\}$ (the ports $\{1, 2, 3, 4\}$ shown in Fig. S1(b)). From $[S_{PPW}]$ we can investigate the mutual coupling one can achieve by exciting the two sources with arbitrary feed networks.

We use unequal Wilkinson power splitters (WPS) shown in Fig. S1(a) to feed the required excitations $\{I_a, I_b\}$ to Sources 1 and 2. The WPS accepts an input current $I_0$ and sends currents $I_a, I_b$ to the two output ports, as shown in the left and right panels of Fig. S1(b). A feature of the WPS is that a resistor decouples Ports 2 and 3. Hence The S-matrix of the WPS is [42]

$$[S_{WPS}] = \begin{bmatrix} 0 & I_a/I_0 & I_b/I_0 \\ I_a/I_0 & 0 & 0 \\ I_b/I_0 & 0 & 0 \end{bmatrix} \qquad (S5)$$

The overall system is thus formed by the proper connection of $[S_{WPS}]$ and $[S_{PPW}]$. Figure S1(a) shows the cable connections (solid lines) and Figure S1(b) shows the signal flow diagram (arrows) of the overall microwave system. From the latter one can deduce the mutual coupling of the overall system as

$$S_{total,21} = W_{21}^2 S_{31} + W_{21}W_{31}(S_{32} + S_{41}) + W_{31}^2 S_{42}. \qquad (S6)$$

where $W_{ij}$ and $S_{ij}$ are the $ij$'th component of $[S_{WPS}]$ and $[S_{PPW}]$ respectively. In our investigation on mutual coupling, we use a 4-port simulation to obtain $[S_{PPW}]$ for each separation distance $d$. Then, using Eq. (S6), we calculate from $[S_{PPW}]$ the mutual coupling for any combination of $\{I_a, I_b\}$. The resulting mutual coupling strengths are plotted Fig. 3 of the main text.

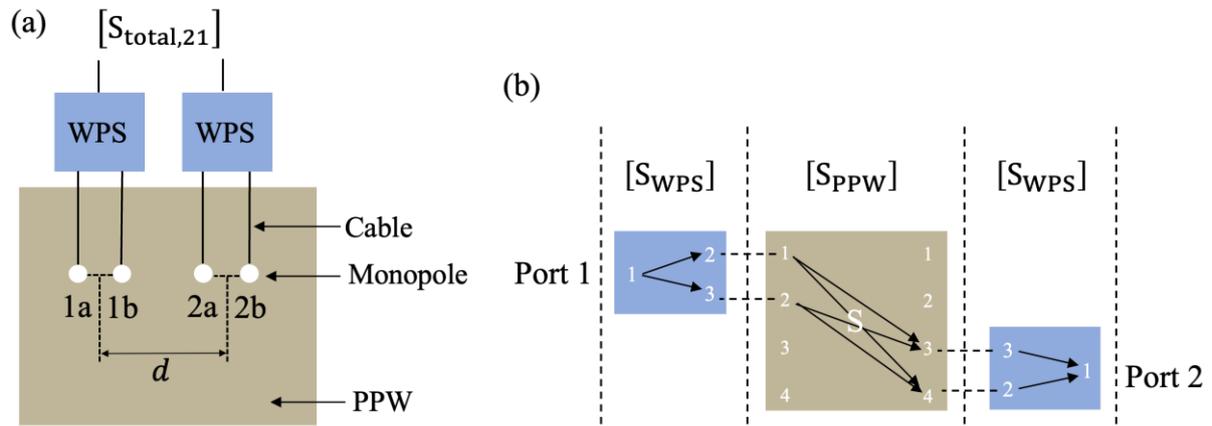

FIG. S1. (a) A schematic of the modelled structure, including the Wilkinson power splitter (WPS), and two TCFs. (b) The corresponding signal flow diagram.

# References

[1] A. Fresnel. *Mémoire sur la diffraction de la lumière*, Paris, Impr. Imperial, **1**, 247363 (1870).

[2] G. Kirchhoff. *Zur Theorie der Lichtstrahlen*. Annalen der Physik (in German). Wiley. **254** (4): 663695 (1883).

[3] A. E. H. Love, *I. The Integration of the Equations of Propagation of Electric Waves*, Philos. Trans. R. Soc. Lond. Ser. Contain. Pap. Math. Phys. Character **197**, 1 (1901).

[4] S. A. Schelkunoff, *Some Equivalence Theorems of Electromagnetics and Their Application to Radiation Problems*, Bell Syst. Tech. J. **15**, 92 (1936).

[5] A. V. Picardi, F. J. Rodríguez-Fortuño, and M. F. Picardi, *Janus and Huygens Dipoles: Near-Field Directionality Beyond Spin-Momentum Locking*, Phys. Rev. Lett. **120**, 117402 (2018).

[6] A. I. Kuznetsov, A. E. Miroshnichenko, M. L. Brongersma, Y. S. Kivshar, and B. Luk'yanchuk, *Optically Resonant Dielectric Nanostructures*, Science **354**, aag2472 (2016).

[7] D. Permyakov et al., *Probing Magnetic and Electric Optical Responses of Silicon Nanoparticles*, Appl. Phys. Lett. **106**, 171110 (2015).

[8] C. Pfeiffer and A. Grbic, *Metamaterial Huygens' Surfaces: Tailoring Wave Fronts with Reflectionless Sheets*, Phys. Rev. Lett. **110**, 197401 (2013).

[9] M. Chen, M. Kim, A. M. H. Wong, and G. V. Eleftheriades, *Huygens' Metasurfaces from Microwaves to Optics: A Review*, Nanophotonics **7**, 1207 (2018).

[10] K. Fan, J. Zhang, X. Liu, G.-F. Zhang, R. D. Averitt, and W. J. Padilla, *Phototunable Dielectric Huygens' Metasurfaces*, Adv. Mater. **30**, 1800278 (2018).

[11] A. M. H. Wong and G. V. Eleftheriades, *Active Huygens' Box: Arbitrary Electromagnetic Wave Generation With an Electronically Controlled Metasurface*, IEEE Trans. Antennas Propag. **69**, 1455 (2021).

[12] K.-M. Luk and H. Wong, *A New Wideband Unidirectional Antenna Element*, Int J Microw Opt Technol **1**, 35 (2006).

[13] L. Ge and K. M. Luk, *A Low-Profile Magneto-Electric Dipole Antenna*, IEEE Trans. Antennas Propag. **60**, 1684 (2012).

[14] R. W. Ziolkowski, *Low Profile, Broadside Radiating, Electrically Small Huygens Source Antennas*, IEEE Access **3**, 2644 (2015).

[15] R. W. Ziolkowski, *Using Huygens Multipole Arrays to Realize Unidirectional Needle-Like Radiation*, Phys. Rev. X **7**, 031017 (2017).

[16] M. F. Picardi, A. V. Zayats, and F. J. Rodríguez-Fortuño, *Not Every Dipole Is the Same: The Hidden Patterns of Dipolar near Fields*, Europhys. News **49**, 14 (2018).

[17] M. F. Picardi, M. Neugebauer, J. S. Eismann, G. Leuchs, P. Banzer, F. J. Rodríguez-Fortuño, and A. V. Zayats, *Experimental Demonstration of Linear and Spinning Janus Dipoles for Polarisation- and Wavelength-Selective near-Field Coupling*, Light Sci. Appl. **8**, 52 (2019).

[18]   Y. Long, H. Ge, D. Zhang, X. Xu, J. Ren, M.-H. Lu, M. Bao, H. Chen, and Y.-F. Chen, *Symmetry Selective Directionality in Near-Field Acoustics*, Natl. Sci. Rev. **7**, 1024 (2020).

[19]   Y. Long, J. Ren, Z. Guo, H. Jiang, Y. Wang, Y. Sun, and H. Chen, *Designing All-Electric Subwavelength Metasources for near-Field Photonic Routings*, Phys. Rev. Lett. **125**, 157401 (2020).

[20]   L. Wei and F. J. Rodríguez-Fortuño, *Far-Field and near-Field Directionality in Acoustic Scattering*, New J. Phys. **22**, 083016 (2020).

[21]   L. Wei and F. J. Rodríguez-Fortuño, *Momentum-Space Geometric Structure of Helical Evanescent Waves and Its Implications on Near-Field Directionality*, Phys. Rev. Appl. **13**, 014008 (2020).

[22]   T. Wu, A. Baron, P. Lalanne, and K. Vynck, *Intrinsic Multipolar Contents of Nanoresonators for Tailored Scattering*, Phys. Rev. A **101**, 011803 (2020).

[23]   Y. Jiang, X. Lin, and and H. Chen, *Directional Polaritonic Excitation of Circular, Huygens and Janus Dipoles in Graphene-Hexagonal Boron Nitride Heterostructures*, Prog. Electromagn. Res. **170**, 169 (2021).

[24]   Y. Zhong, X. Lin, J. Jiang, Y. Yang, G.-G. Liu, H. Xue, T. Low, H. Chen, and B. Zhang, *Toggling Near-Field Directionality via Polarization Control of Surface Waves*, Laser Photonics Rev. **15**, 2000388 (2021).

[25]   Y. Cheng, K. A. Oyesina, B. Xue, D. Lei, A. M. H. Wong, and S. Wang, *Directional Dipole Dice: Circular-Huygens-Janus Dipole in an Anisotropic Chiral Particle*, arXiv:2208.04151.

[26]   H. Jiang, J. Wang, G. Song, J. Ren, X. Yang, J. Xu, and Y. Yang, *Near-Field Properties of Spin, Huygens and Janus Sources in a Narrow Sandwiched Structure*, J. Phys. B At. Mol. Opt. Phys. **55**, 155001 (2022).

[27]   M. F. Picardi, C. P. T. McPolin, J. J. Kingsley-Smith, X. Zhang, S. Xiao, F. J. Rodríguez-Fortuño, and A. V. Zayats, *Integrated Janus Dipole Source for Selective Coupling to Silicon Waveguide Networks*, Appl. Phys. Rev. **9**, 021410 (2022).

[28]   B. Xue and A. M. H. Wong, *Active Janus and Huygens Sources: Achieving Near-Field and Far-Field Directionality Control*, in *2022 IEEE International Symposium on Antennas and Propagation and USNC-URSI Radio Science Meeting (AP-S/URSI)* (2022), pp. 974–975.

[29]   K.-C. Lin, C.-H. Wu, C.-H. Lai, and T.-G. Ma, *Novel Dual-Band Decoupling Network for Two-Element Closely Spaced Array Using Synthesized Microstrip Lines*, IEEE Trans. Antennas Propag. **60**, 5118 (2012).

[30]   X. Tang, X. Qing, and Z. N. Chen, *Simplification and Implementation of Decoupling and Matching Network With Port Pattern-Shaping Capability for Two Closely Spaced Antennas*, IEEE Trans. Antennas Propag. **63**, 3695 (2015).

[31]   S. N. Venkatasubramanian, L. Li, A. Lehtovuori, C. Icheln, and K. Haneda, *Impact of Using Resistive Elements for Wideband Isolation Improvement*, IEEE Trans. Antennas Propag. **65**, 52 (2017).

[32]   F. Amin, R. Saleem, T. Shabbir, S. ur Rehman, M. Bilal, and M. F. Shafique, *A Compact Quad-Element UWB-MIMO Antenna System with Parasitic Decoupling Mechanism*, Appl. Sci. **9**, 11 (2019).

[33] M. Li, L. Jiang, and K. L. Yeung, *A Novel Wideband Decoupling Network for Two Antennas Based on the Wilkinson Power Divider*, IEEE Trans. Antennas Propag. **68**, 5082 (2020).

[34] J.-Y. Lee, S.-H. Kim, and J.-H. Jang, *Reduction of Mutual Coupling in Planar Multiple Antenna by Using 1-D EBG and SRR Structures*, IEEE Trans. Antennas Propag. **63**, 4194 (2015).

[35] Z. Qamar, U. Naeem, S. A. Khan, M. Chongcheawchamnan, and M. F. Shafique, *Mutual Coupling Reduction for High-Performance Densely Packed Patch Antenna Arrays on Finite Substrate*, IEEE Trans. Antennas Propag. **64**, 1653 (2016).

[36] S. Xu, M. Zhang, H. Wen, and J. Wang, *Deep-Subwavelength Decoupling for MIMO Antennas in Mobile Handsets with Singular Medium*, Sci. Rep. **7**, 1 (2017).

[37] L. Sun, Y. Li, Z. Zhang, and H. Wang, *Self-Decoupled MIMO Antenna Pair With Shared Radiator for 5G Smartphones*, IEEE Trans. Antennas Propag. **68**, 3423 (2020).

[38] Q. X. Lai, Y. M. Pan, S. Y. Zheng, and W. J. Yang, *Mutual Coupling Reduction in MIMO Microstrip Patch Array Using TM10 and TM02 Modes*, IEEE Trans. Antennas Propag. **69**, 7562 (2021).

[39] A. M. H. Wong and G. V. Eleftheriades, *A Simple Active Huygens Source for Studying Waveform Synthesis with Huygens Metasurfaces and Antenna Arrays*, in *2015 IEEE International Symposium on Antennas and Propagation & USNC/URSI National Radio Science Meeting* (IEEE, Vancouver, BC, Canada, 2015), pp. 1092–1093.

[40] K. A. Oyesina, *Metasurface-Enabled Cavity Antenna: Beam Steering With Dramatically Reduced Fed Elements*, IEEE ANTENNAS Wirel. Propag. Lett. **19**, 5 (2020).

[41] C. A. Balanis, *Antenna Theory: Analysis and Design* (John wiley & sons, 2015).

[42] D. M. Pozer, *Microwave engineering* (John wiley & sons, 2011).

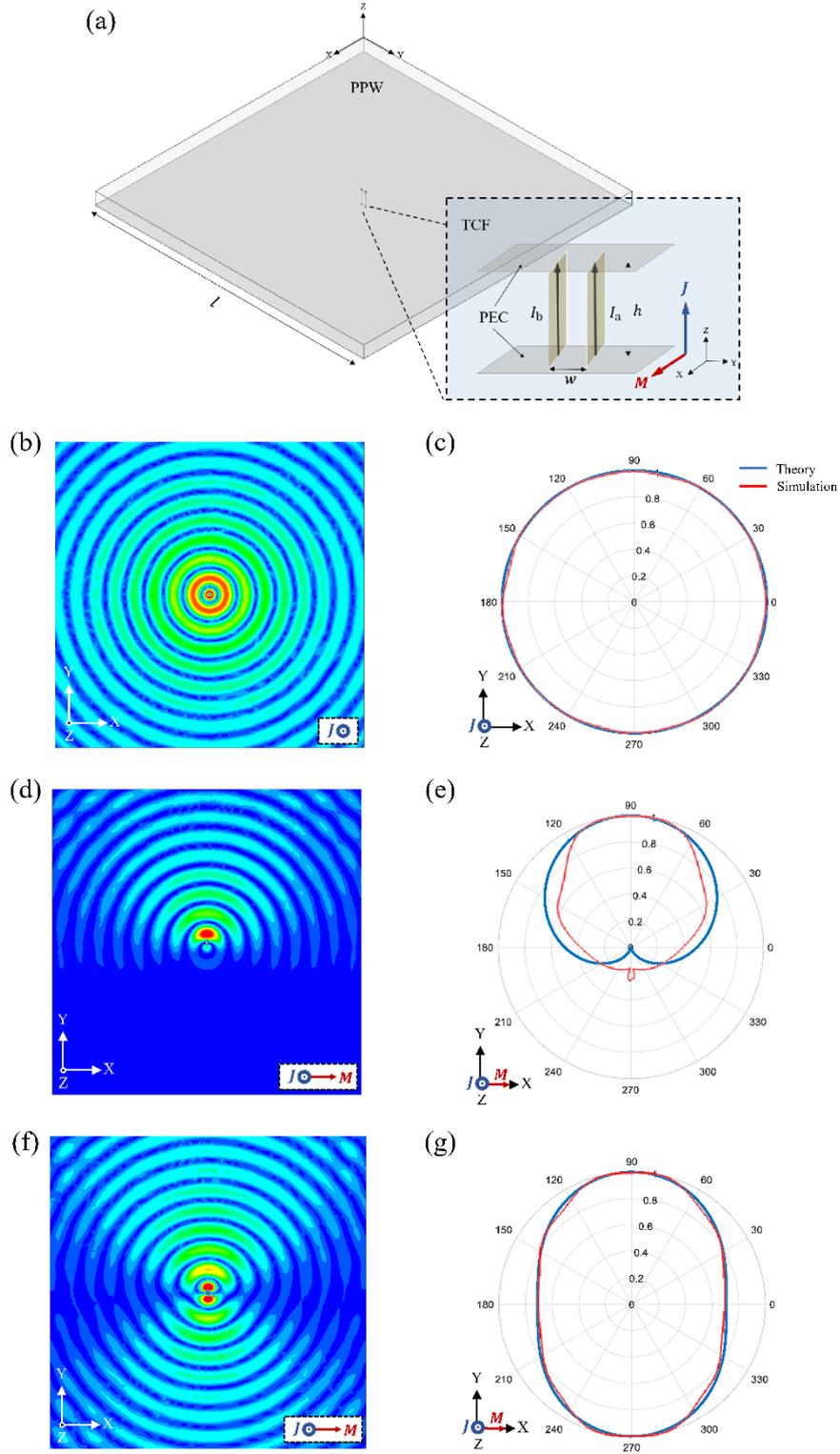

FIG. 1. (a) A diagram of the TCF model within the PPW. h = 45 mm (0.12$\lambda_0$), w = 15 mm (0.04$\lambda_0$) and simulations are performed at 800 MHz. (b, c, d) Simulated |**E**| within the PPW in the area surrounding (b) the electric dipole, (c) the Huygens source and (d) the Janus source. (e, f, g) Calculated 2D radiation patterns of (e) the electric dipole, (f) the Huygens source and (g) the Janus source.

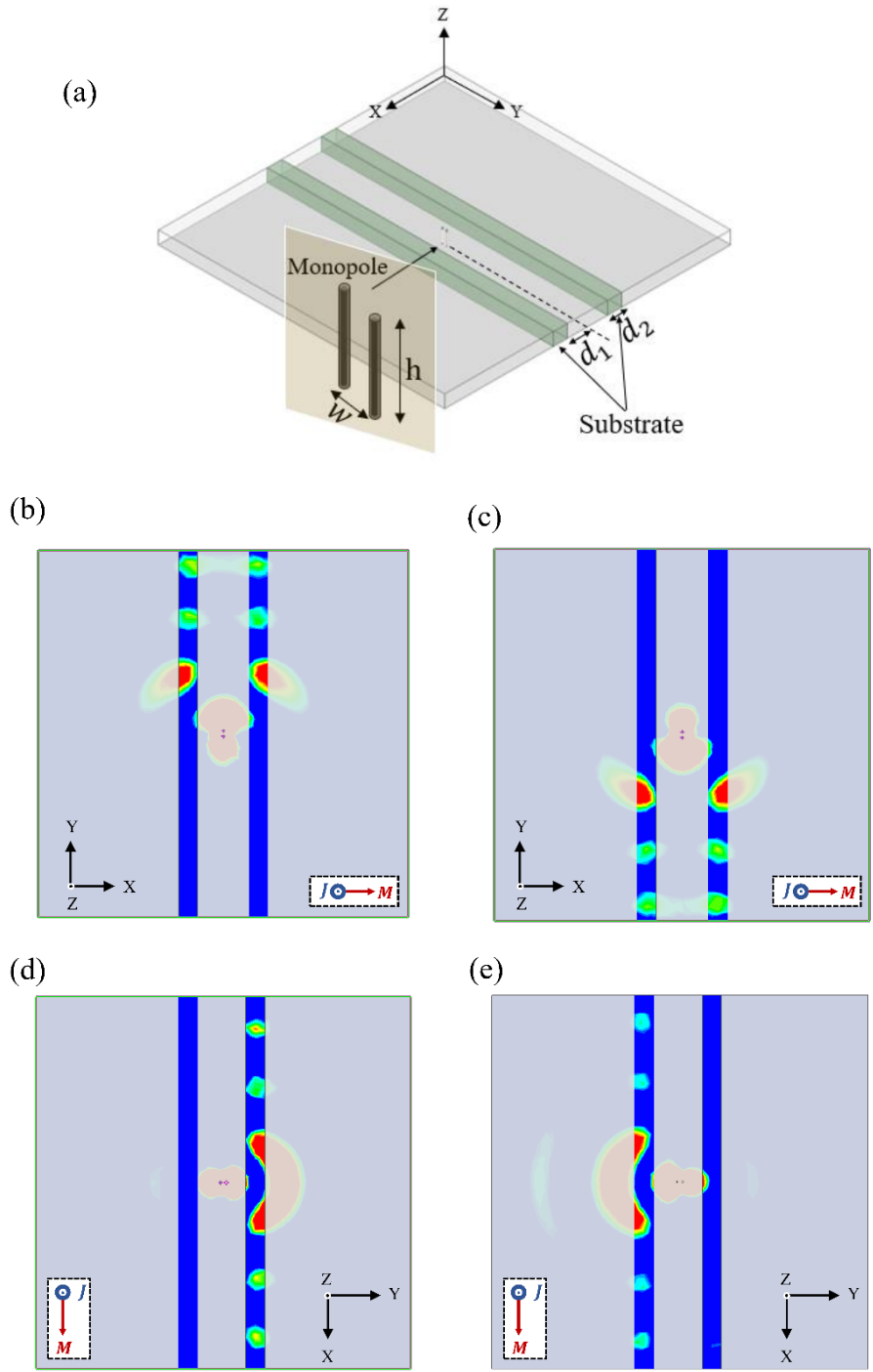

FIG. 2. (a) A diagram of the source in PPW with two substrates. $\varepsilon_r$ (substrate) = 3, $d_1$ = 60 mm, $d_2$ = 50 mm. (b-c) Simulated coupling directionality for the Huygens source, toward the (b) +y and (c) -y directions. (d-e) Simulated coupling directionality for the Janus source, toward the (d) +y and (e) -y directions.

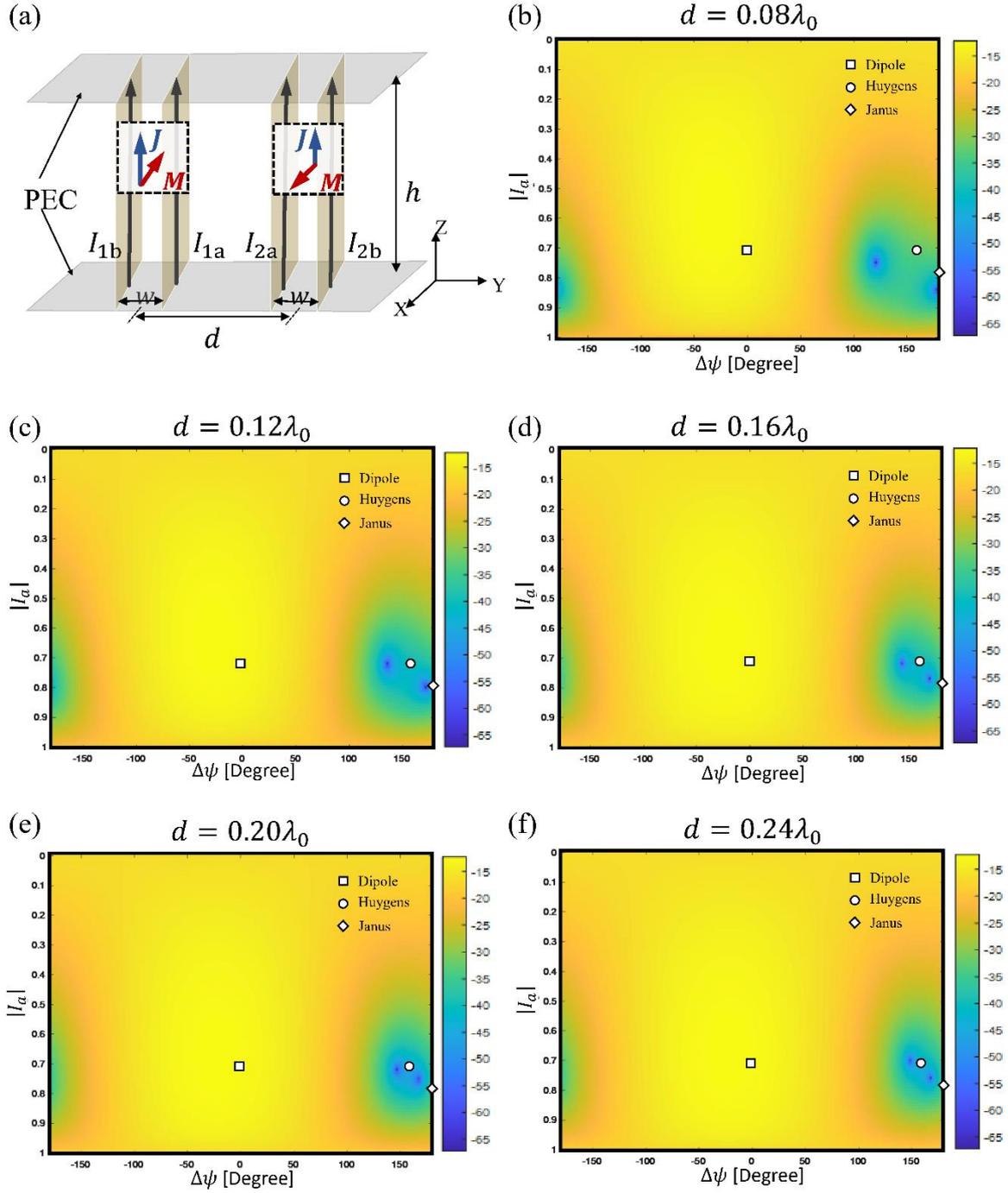

FIG. 3. (a) A diagram of two active sources within the PPW. $h = 0.12\lambda_0$, $w = 0.04\lambda_0$, d is changing from $0.08\lambda_0$ to $0.24\lambda_0$ with $0.04\lambda_0$ interval, and simulations are performed at 800 MHz. 2D coupling coefficient (dB) colour plot of (b) $d = 0.08\lambda_0$, (c) $d = 0.12\lambda_0$, (d) $d = 0.16\lambda_0$, (e) $d = 0.2\lambda_0$, (f) $d = 0.24\lambda_0$.

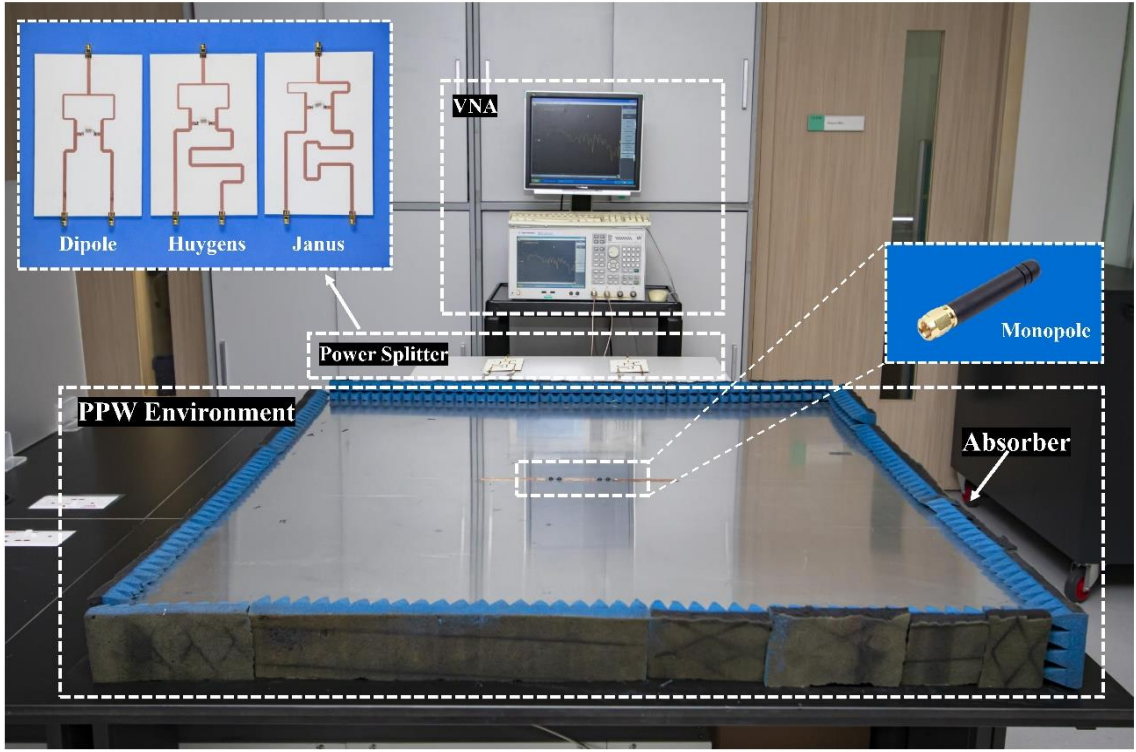

FIG. 4. A photo of the experimental apparatus, showing the PPW environment (monopole used as the practical current), power splitters and the Vector Network Analyzer.

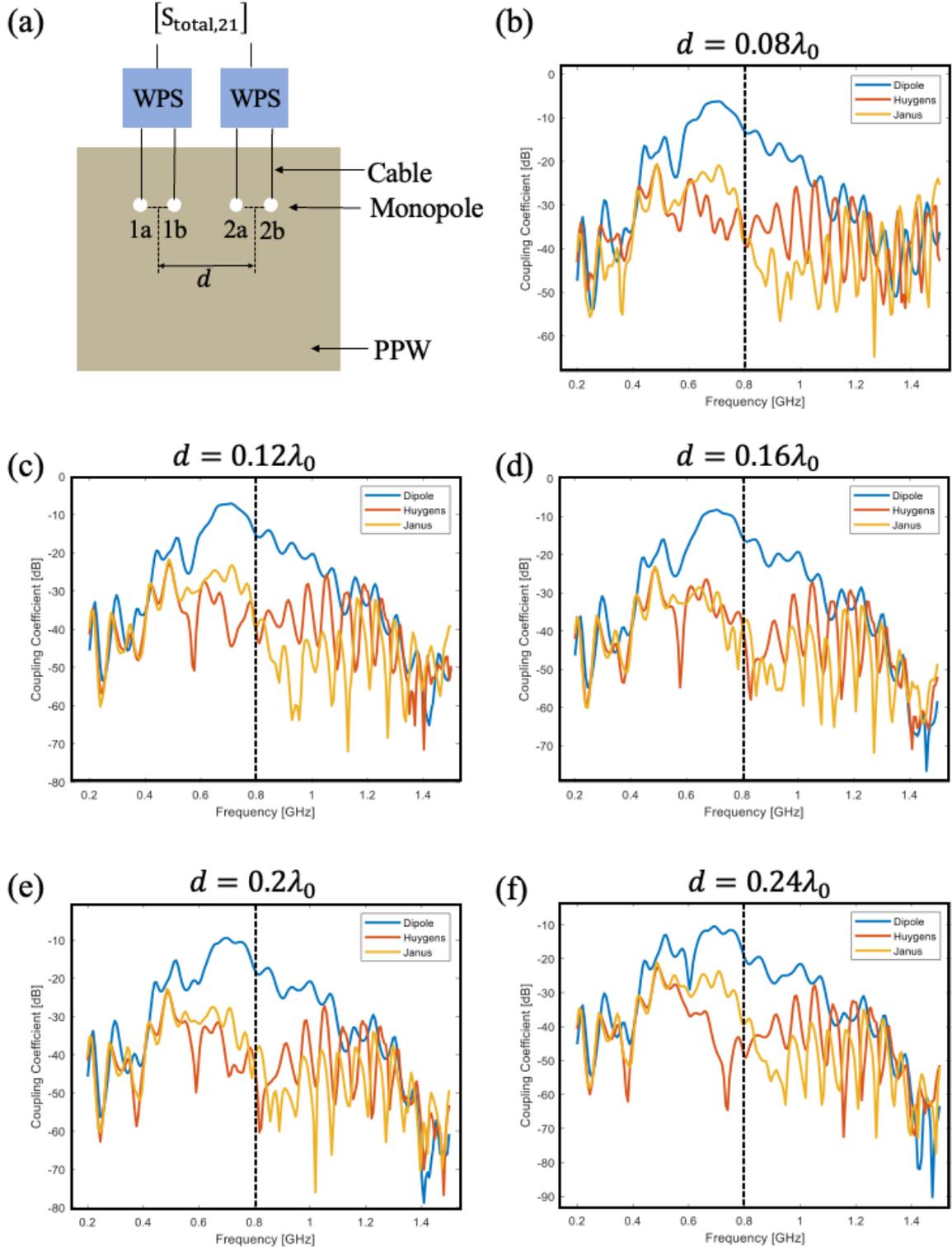

FIG. 5. (a) A schematic of the experiment structure. Experimental mutual coefficient of (b) $d = 0.08\lambda_0$, (c) $d = 0.12\lambda_0$, (d) $d = 0.16\lambda_0$, (e) $d = 0.2\lambda_0$, (f) $d = 0.24\lambda_0$.

TABLE I. Simulated and Measured Coupling Coefficients (dB) of dipole, Huygens source and Janus source with separation d (the distance between two sources) changing from $0.08\lambda_0$ to $0.24\lambda_0$.

|  |  | $d = 0.08\lambda_0$ | $d = 0.12\lambda_0$ | $d = 0.16\lambda_0$ | $d = 0.2\lambda_0$ | $d = 0.24\lambda$ |
|---|---|---|---|---|---|---|
| Dipole | Simulated | -13.08 | -14.81 | -15.95 | -17.01 | -17.66 |
|  | Measured | -12.95 | -15.27 | -15.17 | -17.38 | -19.20 |
| Huygens Source | Simulated | -32.58 | -42.64 | -50.02 | -55.96 | -56.16 |
|  | Measured | -37.41 | -39.33 | -39.21 | -41.47 | -48.53 |
| Janus Source | Simulated | -38.52 | -44.59 | -45.23 | -45.32 | -46.64 |
|  | Measured | -38.63 | -40.14 | -39.74 | -42.45 | -39.51 |